\documentclass[twocolumn,secnumarabic,amssymb, nobibnotes, aps, prl, superscriptaddress]{revtex4-1}
\usepackage{graphicx}
\usepackage[hidelinks,colorlinks=false,linkcolor=blue,urlcolor=green,citecolor=red,]{hyperref} % square links are removed by hidelinks. colorlinks=true changes the square to color links
\usepackage{float} % to use the option H in figure and place exactly in the Latex code position
\usepackage{subfigure}
\usepackage{amsfonts}
\usepackage{amsmath}

\usepackage{xcolor}
\usepackage{color, colortbl}

\usepackage{indentfirst}
\usepackage{graphicx} % Required for inserting images
\usepackage{lineno}
\begin{document}
\title{Sign reversal diode effect in superconducting Dayem nanobridges}
\affiliation{NEST Istituto Nanoscienze-CNR and Scuola Normale Superiore, I-56127, Pisa, Italy}
\affiliation{SPIN-CNR, I-84084 Fisciano (SA), Italy}
\affiliation{Dipartimento di Fisica ``E. R. Caianiello", Universit\`a di Salerno, I-84084 Fisciano (SA), Italy}
\affiliation{present address: CIC nanoGUNE BRTA, E-20018 Donostia-San Sebastián, Spain.}

\author{Daniel Margineda}
\email{d.margineda@nanogune.eu}
\affiliation{NEST Istituto Nanoscienze-CNR and Scuola Normale Superiore, I-56127, Pisa, Italy}

\affiliation{present address: CIC nanoGUNE BRTA, E-20018 Donostia-San Sebastián, Spain.}

\author{Alessandro Crippa}
\affiliation{NEST Istituto Nanoscienze-CNR and Scuola Normale Superiore, I-56127, Pisa, Italy}
\author{Elia Strambini}
\affiliation{NEST Istituto Nanoscienze-CNR and Scuola Normale Superiore, I-56127, Pisa, Italy}
\author{Yuri Fukaya}
\affiliation{SPIN-CNR, I-84084 Fisciano (SA), Italy}
\author{Maria Teresa Mercaldo}
\affiliation{Dipartimento di Fisica ``E. R. Caianiello", Universit\`a di Salerno, I-84084 Fisciano (SA), Italy}
\author{Mario Cuoco}
\affiliation{SPIN-CNR, I-84084 Fisciano (SA), Italy}
\author{Francesco Giazotto}
\email{francesco.giazotto@sns.it}
\affiliation{NEST Istituto Nanoscienze-CNR and Scuola Normale Superiore, I-56127, Pisa, Italy}

%\begin{abstract}
%\end{abstract}

\keywords{Superconducting Diode Effect, Superconducting electronics}
\maketitle
%\section{Abstract}
\noindent\large{\textbf{Abstract}}\normalsize\\
Supercurrent diodes are nonreciprocal electronic elements whose switching current depends on their flow direction. 
Recently, a variety of composite systems combining different materials and engineered asymmetric superconducting devices have been proposed.
Yet, ease of fabrication and tunable sign of supercurrent rectification joined to large efficiency have not been assessed in a single platform so far.
We demonstrate that all-metallic superconducting Dayem nanobridges naturally exhibit nonreciprocal supercurrents under an external magnetic field, with a rectification efficiency up to $\sim 27\%$. 
Our niobium nanostructures are tailored so that the diode polarity can be tuned by varying the amplitude of an out-of-plane magnetic field or the temperature in a regime without magnetic screening. We show that sign reversal of the diode effect 
may arise from the high-harmonic content of the current phase relation in combination with vortex phase windings present in the bridge or an anomalous phase shift compatible with anisotropic spin-orbit interactions.\\

%\section{Introduction}
\noindent\large{\textbf{Introduction}}\normalsize\\
Non-reciprocal charge transport is an essential element in modern electronics as a building block for multiple components such as rectifiers, photodetectors, and logic circuits. 
For instance, pn-junctions and Schottky-barrier devices are archetypal semiconductor-based examples of systems, known as \textit{diodes}, with direction-selective charge propagation. Their operation stems from the spatial asymmetry of the heterojunction that provides inversion symmetry breaking. 
Likewise, dissipationless rectification refers to the asymmetric switching of the critical current $(I_{sw})$  required to turn a superconductor into the normal state depending on the current bias polarity.  
Breaking both inversion and time-reversal symmetry, which are preserved in conventional superconductors, is the foundational aspect to enable the diode effect, as recently observed in superconducting materials~\cite{wak17,mer23,lin22,paolucci2023gate} and  heterostructures~\cite{and20,pal22,baur22,bau22,jeo22, pal22,sun23,lot23,tur22}.
Recent experimental findings have boosted a number of theoretical investigations 
in superconductors~\cite{dai22,ber22,yua22,he22} and Josephson junctions (JJs)~\cite{mis21,dav22,zha22}. 
In particular, several mechanisms have been proposed to account for the supercurrent diode effect (SDE). On the one hand, those based on intrinsic depairing currents focus on finite momentum pairing that arise from the combination of spin-orbit coupling (SOC) and Zeeman field~\cite{dai22,ber22,yua22,he22,Scammell_2022}, or from Meissner currents~\cite{dav22}. 
On the other hand, other works underline the role of Abrikosov vortices, magnetic fluxes, screening currents or self-field effects as key elements for setting out non-reciprocal charge transport in superconductors~\cite{wambaugh99,vodolazov05,villegas05,devondel05,car09,cerbu_2013}, such as in systems with trapped Abrikosov vortices~\cite{gol21,lyu21,gol22} or in micron-sized Nb-based strips~\cite{sur22,cha23,hou22,sat23,vav13}.

Till now, most of the research efforts have aimed to the realization of a SDE that maximizes the rectification efficiency, while the change of its polarity has been reported in a few cases only~\cite{lot23,pal22,cos22,sun23,kawarazaki2022,Gupta2023}. The SDE sign reversal has been interpreted as a consequence of
finite momentum pairing~\cite{pal22,cos22,kawarazaki2022,lot23} requiring in-plane magnetic fields or diamagnetic currents and Josephson vortices~\cite{gol22,sun23}, as well as ascribed to vortex ratchet and asymmetric pinning effects~\cite{Gillijns07,He19,Ideue20,Ji21}. All these outcomes suggest the need for an effective mastering of the polarity change of the SDE and 
its implementation in a simple and monolithic platform suitable for nanoscale miniaturization, not
accomplished yet.
\begin{figure*}[ht!]
\centering
\includegraphics[scale=0.18]{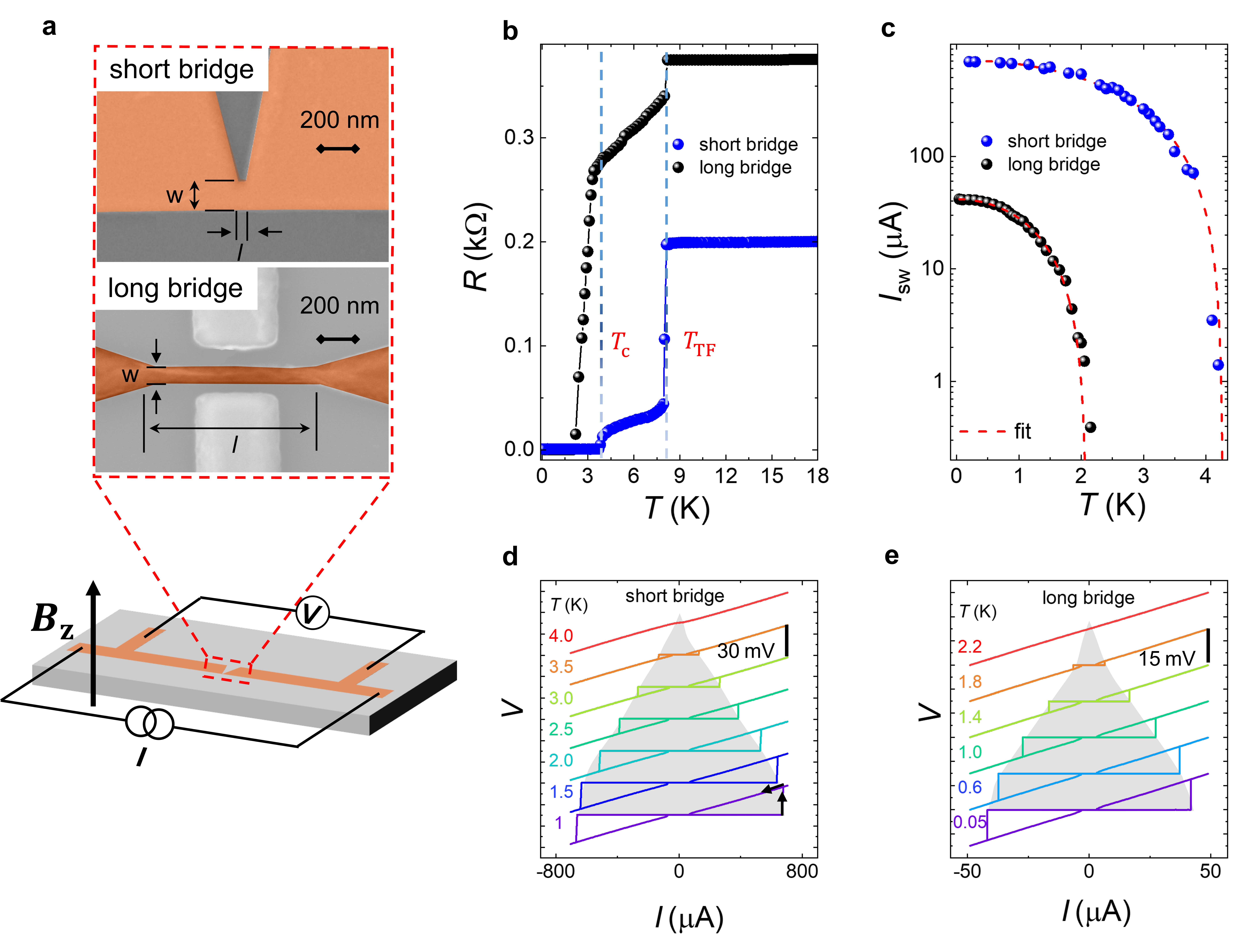}
\caption{\textbf{Nb Dayem nanobridge diodes and basic electrical characterization.} \textbf{a}, Samples and schematic setup to measure the voltage characteristics $V$ as a function of a bias current $I$ with an applied out-of-plane magnetic field $B_z$. In the upper part, scanning electron micrographs of two weak links with different length $l$ and width $w$: one is a  constriction of Nb strip (with characteristic dimensions $l\sim80$\,nm and $w\sim180$\,nm), the other is a quasi-1D wire ($l\sim1$\,$\mu$m, $w\sim80$\,nm) connecting the banks; they are labeled as short and long, respectively. 
The electrodes used in the experiment are false-colored in orange. \textbf{b}, Temperature dependence of the zero-bias resistance for the two Dayem bridges. Thin film $(T_{TF})$ and weak link $(T_c)$ critical temperatures are marked by dashed lines for the short device. 
\textbf{c}, Temperature $(T)$ dependence of the switching supercurrents for the short (blue dots) and the long (black dots) bridges. Red dashed lines are the fit to Bardeen equation, as described in the text. 
$IV$ characteristics of the short, \textbf{d}, and  long bridge, \textbf{e}. The curves are vertically offset for clarity, and the superconducting region is highlighted in grey to visualize the temperature-induced decay of the dissipationless current. Black arrows indicate the direction of the bias current swept back and forth starting at zero amplitude.}
\label{samples} 
\end{figure*}
%\vspace{-0.1cm}

Here, we experimentally demonstrate a sign reversal tunable SDE in elemental superconducting weak links made on niobium (Nb). Nano-sized constrictions of Nb realize Dayem bridges whose switching currents for positive and negative sweep direction, $I^+_{sw}$ and $I^-_{sw}$, respectively, differ in the absolute value. This difference can be tuned both in amplitude and sign by an out-of-plane magnetic field $(B_z)$,  without inverting the polarity of $B_z$.
Thermal effects can lead to two different energy scales for the maximal amplitude and the sign reversal of the diode efficiency which is inconsistent with physical scenarios related to Meissner currents or self-field effects. 
We show that sign reversal of the non-reciprocal response may arise from the phase shift due to the vortex phase winding or from spin-orbit effects due to the material granularity, in either case jointly with a few-harmonic content of the current-phase relation (CPR) of the weak link.\\

\begin{figure*}[ht]
\centering
\includegraphics[scale=0.16 ]{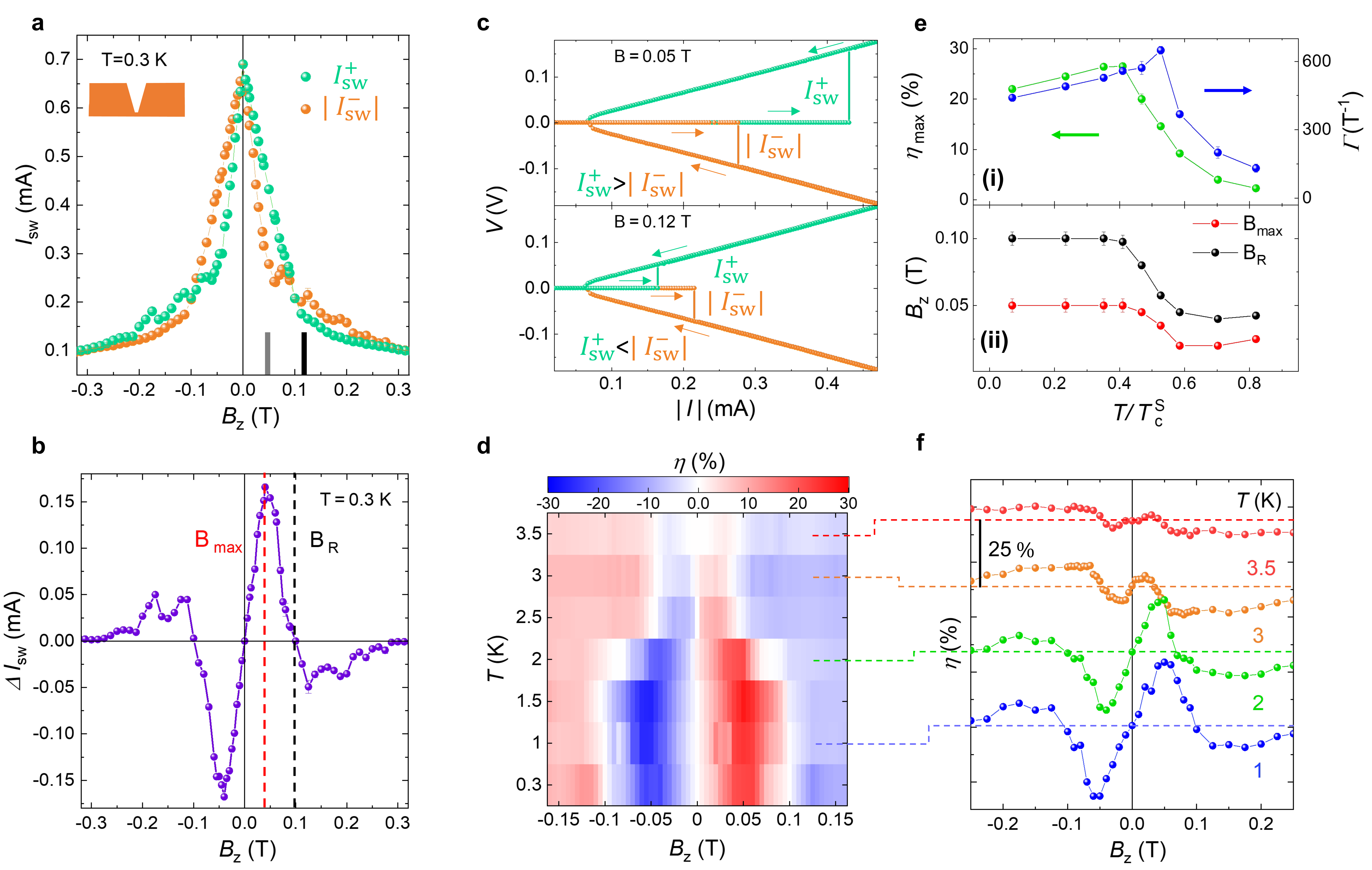}
\caption{\textbf{Diode effect in a short Dayem bridge.} \textbf{a}, Out-of-plane magnetic field dependence of the switching current for positive $I_{sw}^+$(green dots) and negative $|I_{sw}^-|$(orange dots) bias current recorded at 300\,mK. 
\textbf{b}, $\Delta I_{sw}$ obtained from \textbf{a}. 
The magnetic field values at which the maximum $(B_{max})$ and sign reversal $(B_R)$ of the rectification occur are marked by red and black dashed lines, respectively. 
\textbf{c}, $IV$ curves with positive $I_{sw}^+>|I_{sw}^-|$ and negative $I_{sw}^+<|I_{sw}^-|$ rectification recorded at the magnetic fields marked by bars in panel \textbf{a}. 
\textbf{d}, Color plot of the rectification efficiency $\eta(B_z,T)$ versus bath temperature and magnetic field. 
\textbf{e}, Rectification parameters: $\eta_{max} \equiv (\eta_{max}(B_z>0)+\eta_{max}(B_z<0))/2$ (green dots), field-to-rectification efficiency transfer function $\Gamma$ (blue dots) \textbf{(i)}, $B_{max}$ (red dots) and $B_R$ (black dots) magnetic fields \textbf{(ii)} versus normalized temperature. $T_c^S$ denotes the critical temperature of the short bridge. $\eta_{max}$ is the rectification value of the low-field peak at $B_{max}$. \textbf{f}, $\eta (B_z)$ for selected values of bath temperature marked by dashed lines in panel \textbf{d}.
Curves are vertically offset for clarity. $\eta(B_z)$ function exhibits two extrema, below and above $B_{R}$. The rectification peak $\eta_{max}$, identified by $B_{max}$, decreases in magnitude and field for $T\geq$ 1.75\,K = 0.4 $T_c^S$, until the minimum in the rectification at $B>B_R$ becomes an absolute extremum at T = 3\,K (orange dots). For discussion, we keep $B_{max}$ and $\eta_{max}$ as the nomenclature for the maximum rectification.}
\label{shortNB} 
	\end{figure*}

\noindent\large{\textbf{Results}}\normalsize\\
\textbf{Metallic diode architectures.}	
We analyze two different geometries of Nb Dayem bridges, i.e., weak links made of a constant-thickness and all-metallic constriction between two superconducting banks~\cite{lik79}. The schematics of the electronic circuitry and false-color scanning electron micrographs of the devices are shown in Fig.~\ref{samples}a.
In the first type of samples, 25-nm-thick micrometer-wide banks are connected via a link whose length $l$ is $\sim 80$ nm and width $w\sim 180$ nm. 
The second type consists of 55-nm-thick banks connected via a quasi-one-dimensional wire with $l\simeq1$\,$\mu$m, and $w\simeq80$\,nm. 
Hereafter, we shall refer to the first and second types of bridges as short and long, respectively. 
Both device families are patterned through a single electron-beam lithography step followed by sputter deposition of the Nb thin film and lift-off. A 4-nm-thick Ti layer is pre-sputtered for adhesion purposes.

The differential resistance $R=dV/dI$ versus temperature $T$ of two representative bridges is shown in Fig.~\ref{samples}b.
The first abrupt reduction of $R$ marks the critical temperature of the Nb films $T_{TF} \simeq 8.1(7.9)$ K for the 55(25)-nm-thick sample, suggesting minor impact of the film thickness on the superconducting state.
The resistance drops to zero at the critical temperature of the weak link $(T_c)$, which strongly depends, along with its normal-state resistance $R_N$, on the geometry~\cite{lik79}. While the short bridge exhibits $R_N \sim$ 40\,$\Omega$, in the long one has $R_N \sim$ 270\,$\Omega$.

Below $T_c$, dissipationless transport occurs in the bridges owing to Cooper pairs supercurrent. 
The temperature dependence of the switching current $I_{sw}$ of both devices is displayed in Fig~\ref{samples}c. 
From the fit to the Bardeen equation $I_{sw}(T) =I_{sw}^0[1-(\frac{T}{T_{c}})^2]^{\frac{3}{2}}$, 
we extract a zero-temperature switching current $I_{sw}^0\simeq 720$ $\mu$A and a critical temperature $T_c^S\simeq 4.3$\,K for the shor bridge. 
Similarly, for the long weak link we obtain $I_{sw}^0\simeq 42\,\mu$A and $T_c^L\simeq 2.1$\,K. 
From these values, we determine a zero-temperature BCS energy gap $\Delta_0$=1.764 $k_B T_{c}^{S (L)}\simeq$ 650(320)\,$\mu$eV for the short(long) bridge, where $k_B$ is the Boltzmann constant. 
For the long bridge, we deduce a superconducting coherence length $\xi_0=\sqrt{\hbar l/(R_N wt e^2 N_F \Delta_0)}\simeq$ 11\,nm, where $t$ is the film thickness, $N_F \simeq 5.33\times 10^{47} J^{-1}m^{-3}$ is the density of states at the Fermi level of Nb~\cite{jan88}, and $e$ is the electron charge. 
Similarly, we can evaluate the London penetration depth $\lambda_L=\sqrt{\hbar R_N wt/( \pi l \mu_0  \Delta_0)}\simeq$ 790\,nm, where $\mu_0$ is the vacuum magnetic permeability. 
Since $w,t \ll \lambda_L$, the bridges can be uniformly penetrated by an external magnetic field.

The current vs voltage $(IV)$ characteristics of the short and long bridges are shown in Fig.\ref{samples}d and e, respectively, for selected values of bath temperature. 
The devices show an abrupt transition to the normal state at the switching current $I_{sw}$ and display the typical hysteresis of metallic junctions which originates from Joule heating induced in the bridge when the bias current is swept back from the resistive to the dissipationless state~\cite{sko76}.\\ 
%%%%%%%%%%%%%%%%   fig 3 %%%%%%%%%%%%%%%%%%%%%%%%
\begin{figure*}[ht!]
\includegraphics[scale=0.16]{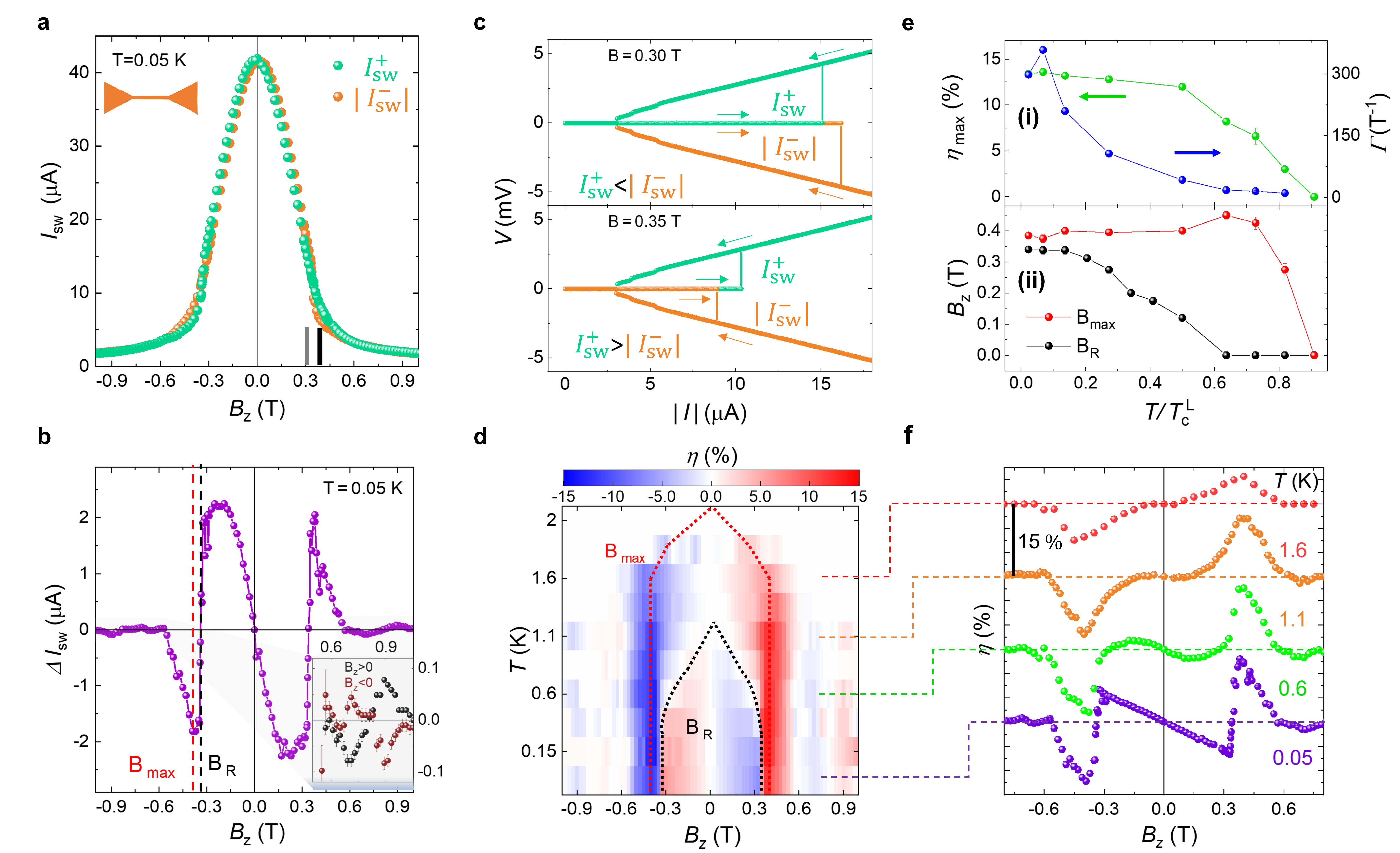}
\caption{\textbf{Diode effect in a long Dayem bridge.} \textbf{a}, Magnetic field dependence of the switching current for positive $I_{sw}^+$(green dots) and negative $|I_{sw}^-|$(orange dots) bias current recorded at 50\,mK. \textbf{b}, $\Delta I_{sw}$ obtained from panel \textbf{a}.  
The rectification efficiency increases linearly in $B_z$ until $|B_z|\simeq 015$\,T. 
Red and black dashed lines mark the magnetic field values corresponding to sign reversal $(B_R)$  and maximum rectification $(B_{max})$, respectively. 
Inset: Blow-up of $\Delta I_{sw}$ for large positive (black) and negative (brown) fields displaying several changes of signs. \textbf{c}, $IV$ characteristics with negative $I_{sw}^+<|I_{sw}^-|$ and positive $I_{sw}^+>|I_{sw}^-|$ rectification recorded at the magnetic fields marked by bars in panel \textbf{a}.
\textbf{d}, Color plot of the rectification efficiency as a function of temperature and magnetic field, $\eta(B_z,T)$. Dashed lines are guides for the eye to highlight the different temperature trends in $B_{max}$ and $B_R$. 
\textbf{e}, Rectification parameters: $\eta_{max} \equiv (\eta_{max}(B_z>0)+\eta_{max}(B_z<0))/2$ (green dots), field-to-rectification efficiency transfer function $\Gamma$ (blue dots) \textbf{(i)}, $B_{max}$ (red dots) and $B_R$ (black dots) magnetic fields \textbf{(ii)} versus normalized temperature. $T_c^L$ denotes the critical temperature of the long bridge. $\eta_{max}$ is the rectification value of the low-field peak at $B_{max}$. \textbf{f} $\eta_{B_z}$ for selected bath temperatures marked by dashed lines in panel \textbf{d}. 
Curves are vertically offset for clarity.}
\label{longNB} 
\end{figure*}
%\section{Short Dayem bridge diode performance}

\noindent\textbf{Short Dayem bridge diode performance.}
Let us now discuss how the short Dayem bridge in Fig.\ref{samples}a can be used as a supercurrent diode. 
Non-reciprocal dissipationless transport is revealed by comparing the switching currents while sweeping the biasing current from zero to positive values ($I_{sw}^+$) or from zero to negative values ($I_{sw}^-$) in the presence of an out-of-plane magnetic field $B_z$. 
The switching currents at $T=0.3$\,K are reported in Fig.~\ref{shortNB}a. The magnetic field increasingly reduces the superconducting gap and thereby both the switching currents. 
A linear decrease in $B_z$ of $I_{sw}^+$ and $|I_{sw}^-|$ is observed up to $\sim0.07$\,T. 
At larger fields, the dependence of the switching currents on $B_z$ is sublinear. Notably, both $I_{sw}(B_z)$ are not antisymmetric with respect to the magnetic field ($I_{sw}(B_z)\neq I_{sw}(-B_z)$) while the symmetry relation $I_{sw}^+(B_z) \simeq -I_{sw}^-(-B_z)$ is respected, within the small experimental fluctuations, as theoretically expected.
This symmetry relation is further confirmed in the switching currents difference $\Delta I_{sw}\equiv I_{sw}^+-|I_{sw}^-|$ displaying an odd-in-$B_z$ superconducting diode effect ($\Delta I_{sw}(B_z)\simeq-\Delta I_{sw}(-B_z)$) as shown in Figure~\ref{shortNB}b. $\Delta I_{sw}$ is characterized by a maximum at $B_{max}\simeq 0.05$\,T and a sign inversion at $B_R \simeq  0.1$\,T where $I_{sw}^+$ and $|I_{sw}^-|$ have a crossing (see Fig.~\ref{shortNB}a). From now on, $B_{max}$ indicates the position in field of the rectification peak.
Two $IV$ curves, recorded for magnetic fields lower and larger than $B_R$, are plotted in Fig.~\ref{shortNB}c to emphasize the sign change in the rectification.

Nonreciprocal transport can be conveniently quantified by the rectification efficiency defined as $\eta=\frac{I_{sw}^+-|I_{sw}^-|}{I_{sw}^++|I_{sw}^-|}$. 
Figure~\ref{shortNB}d shows the evolution of $\eta$ versus $B_z$ and $T$. $\eta(B_z)$ is substantially unaffected by thermal effects up to $T\simeq 1.75$\,K $= 0.41\,T_{c}^S$ where  a maximum rectification $\eta_{max}\sim 27\%$ is obtained. The evolution of $\eta_{max}$ in temperature is displayed in % the top panel of Fig.~\ref{shortNB}e (left vertical axis).
Fig.~\ref{shortNB}e\,(i).
In addition, we parametrize the diode sensitivity to the magnetic field in the vicinity of the abrupt sign change as $\Gamma=\eta_{max}/(|B_{max}-B_R|)$. 
A maximum value $\Gamma \sim 650$\,T$^{-1}$ is achieved around $2.25$\,K (see Fig.~\ref{shortNB}e\,(i)) %(see Fig.~\ref{shortNB}e, top panel and right vertical axis).
At higher temperatures, the quantities $\eta_{max}$, $\Gamma$, and the characteristic magnetic fields $B_{max}$ and $B_R$ related to rectification %(see Fig.~\ref{shortNB}e, bottom panel)
(see Fig.~\ref{shortNB}e\,(ii)),
all decrease in a similar fashion.
The full profile of the rectification efficiency versus $B_z$ is better visualized in Fig.~\ref{shortNB}f where $\eta (B_z)$ is plotted for a few selected values of temperature.\\
%\section{Long Dayem bridge diode performance}

\noindent\textbf{Long Dayem bridge diode performance.}
Next, we characterize the long nanobridge shown in Fig.~\ref{samples}a.
Figure~\ref{longNB}a reports the decay of $I_{sw}^+$ and $|I_{sw}^-|$ as a function of $B_z$. 
At first, we notice that the switching currents are damped down to $\sim 60$\% of their zero-field value at $B_z\simeq 0.3$\,T, whereas in the previous sample, the same damping is achieved for lower fields ($B_z\simeq 0.07$\,T see Fig.~\ref{shortNB}a).  
Figure~\ref{longNB}b displays $\Delta I_{sw}$ versus $B_z$. For low magnetic fields, $\Delta I_{sw} (B_z)$ exhibits a linear relation.
While increasing $|B_z|$ further, $\Delta I_{sw}$ bends and then inverts its trend: an abrupt jump realizes a sign reversal at $|B_R| \sim 0.34$\,T. Then, a relative peak at $|B_{max}| \simeq 0.38$\,T, marked by a red dashed line, represents the field at which maximum rectification efficiency is achieved as before. Next to the sign change, $\Delta I_{sw}$ leaves the clean trend and looks noisy. 
Such small jumps are reproducible, thus ruling out a stochastic nature of the underlying processes. 
Finally, $\Delta I_{sw}$ oscillates at higher magnetic fields, as shown in the inset of Fig.~\ref{longNB}b. 
Two $IV$ curves, for fields lower and larger than $B_R$, are plotted in Fig.~\ref{longNB}c to highlight that the rectification sign changes from negative to positive as the field $B_z>0$ increases contrary to the short bridge. This change in symmetry is attributed to the vortex nucleation as discussed later.

The magnetic field and temperature dependence of the rectification efficiency $\eta$ is presented as a color plot in Fig.~\ref{longNB}d. 
The sign change and the maximum rectification are affected by temperature in a different way as compared to the short constriction. 
The linear increase of the rectification at low fields smears out with temperature, reducing $B_R$ until it vanishes at $T \simeq 1.1$\,K $= 0.5\,T_c^L$. Rectification parameters displayed in Fig.~\ref{longNB}e shows that $B_{max}$ is more robust in temperature than $B_R$: it is still observable at $T \simeq 1.8$\,K $= 0.8\,T_c^L$. 
The sudden change of sign is quantified by a maximum $\Gamma \sim 360$\,T$^{-1}$ at $0.15$\,K. 
As before, the profile of rectification efficiency as a function of $B_z$ is shown in Fig.~\ref{longNB}f for a few selected values of temperature.
The difference in the temperature trend between $B_R$ and $B_{max}$ (see Fig.~\ref{longNB}e\,(ii)) suggests two different energy scales responsible for the sign reversal and the maximum rectification, as confirmed by measurements obtained in other devices with same nominal dimensions. (see Supplementary Figure\,1).
Rectification on the second sample exhibits similar $\eta(B_z)$ lineshape with almost identical $B_{max}$ and $\eta_{max}$ values and temperature dependence. In this sample, low-field features fade more rapidly with temperature, which appears to be sample-dependent.\\

\begin{figure*}[ht!]
\includegraphics[scale=0.16]{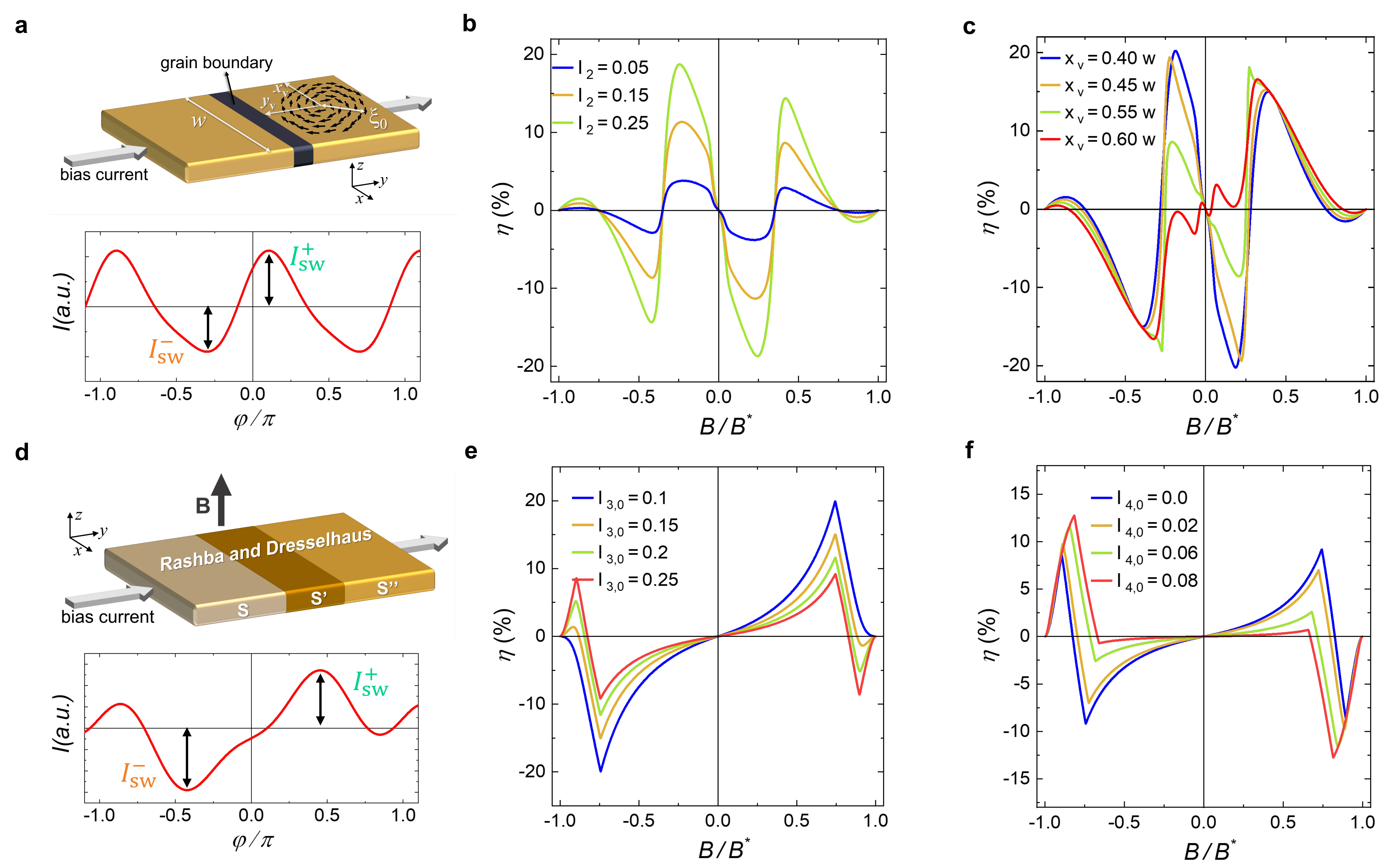}
\caption{\textbf{Modelling supercurrent across a nanobridge for sign-tunable diode effect}. 
\textbf{a}, Sketch of the theoretical framework for model I, with the arrows indicating the phase winding associated with a vortex nucleated near the grain boundary. Characteristic CPR (red line) of the weak link hosting a vortex close to the grain boundary. \textbf{b}, Non-reciprocal rectification efficiency $\eta$ calculated for a few values of the second harmonic at a given position of the vortex core with $(x_v,y_v)=(0.4\,w,0.4\,w)$ and $\gamma_v=1$.
\textbf{c}, Non-reciprocal rectification efficiency $\eta$ for different positions $(x_v)$ of the vortex with $y_v=0.4w$ and $I_{2,0}=0.2$. Variation of the vortex position leads to substantial modifications in $\eta$ at low magnetic fields.
\textbf{d}, Sketch of the nanobridge, with SS'S indicating the regions with different amplitudes of the superconducting gap. Here we assume the presence of Rashba and Dresselhaus spin-orbit couplings. Representative skewed and asymmetric CPR (red line) originated by high-harmonic components (up to the third one) and an anomalous phase offset $\varphi_0$.
\textbf{e}, Non-reciprocal rectification efficiency $\eta$ calculated for several values of the third harmonic component $I_{3,0}$, assuming that $I_{1,0}=1$, $I_{2,0}=-0.3$ and $\Gamma_B=0.2 B/B^*$. $\eta$ changes sign after maximum rectification is reached.
\textbf{f}, Impact of the fourth harmonic in the rectification, with $I_{1,0}=1$, $I_{2,0}=-0.3$ and  $I_{3,0}=0.25$.}
\label{highharmonics} 
\end{figure*}

\noindent\textbf{Modeling the sign reversal of the diode effect.}
Previous SDE reported in conventional superconductor are normally attributed to Abrikosov vortices and self-field effects~\cite{vav13,gol22} or Meissner currents~\cite{sur22,cha23,hou22,sat23}. We have neglected those mechanisms since they are weak in amplitude. Such conclusion can be directly deduced from the stark contrast of the profile of the switching current vs B-field data points. In both scenarios $I_{sw}(B_z)$ reaches a maximum at nonzero magnetic fields. Moreover, amplitude asymmetry does not change sign as a function of the magnetic field without reversing its orientation before the critical current vanishes. Instead, the switching current-field profile in our devices decays monotonically at low fields with no signature of the characteristic skewness. For this reason, we propose two physical scenarios compatible with our devices that may explain our experimental findings. Both of them rely on non-sinusoidal CPRs, typical of superconducting nanobridges~\cite{lik79}, combined with an inversion-symmetry breaker. In model I, this is represented by a supercurrent vortex, while in model II by spin-orbit couplings.
An out-of-plane magnetic field parametrized by $B^*$ is considered, where $B^*$ is given by the zero rectification field $\eta(B^*)=0$ and by the vanishing of high-harmonic amplitude.

In model I, the Dayem nanobridge is schematized as a one-dimensional chain of weak links of width $w$ formed by the Nb grains. A supercurrent vortex can nucleate in one of these weak links~\cite{Aranson1994,Aranson_1996}, as sketched in Fig.~\ref{highharmonics}a, which induces a phase winding in the superconducting order parameter. 
These vortices have a typical size of the order of $\xi$, so only a few of them can be accommodated within the bridge.
Notice that such vorticity is not a screening of the $B$-field, since the small dimensions of the bridge ($w\ll\lambda_L$) allow full penetration of $B_z$.
In this framework, the CPR is affected by two phase shifts: the conventional vector potential associated with $B_z$ and the phase winding of the vortex.
It is indeed the interplay of these two contributions that is responsible for a sign change in the rectification parameter.
Though on a different length scale, this physical scenario is similar to that of Josephson phase vortices~\cite{krasnov20,gol22}. 
The rectification parameter $\eta$ is then evaluated by determining the maximum and minimum values of the Josephson current with respect to the phase bias, see Methods for details.

Figure~\ref{highharmonics}b, reports the evolution of $\eta$ in $B$ for different amplitudes of the second harmonic, $I_2$, of the CPR. The magnitude of $\eta$ scales with $I_2$ showing multiple nodes, whose position in $B$ is independent of $I_2$.
The sign change also depends on the position of the vortex, as displayed in Fig.~\ref{highharmonics}c, where $\eta(B)$ is evaluated for a vortex nucleated at different distances from the lateral edge of the bridge ($x_\nu$).
This behavior suggests a phase-shift competition dominated at low fields by the vortex phase slip and at large fields by the vector potential.

Another scenario able to describe the sign reversal in the diode rectification can be envisioned by combining the colored CPR of the nanobridge with an anomalous phase shift~\cite{ass19,str20,may20,mar23,ber22} induced by spin-orbit interactions and magnetic fields. Nb is a heavy metal and thus possesses an atomistic spin-orbit interaction of the order of $\sim 100$ meV~\cite{mac69}.
We extend such concept to our weak links where we call for a spin-orbit coupling of the order of a few meV, i.e., 2 orders of
magnitude weaker than the atomic limit~\cite{gul22}.
In particular, we can expect that mirror symmetry can be locally or globally broken in polycrystalline films~\cite{Park19}, thereby leading to spin-orbit interaction of both Rashba and Dresselhaus types (see Methods). 
Figure~\ref{highharmonics}d shows a sketch of the bridge modeled as an effective $SS'S$ structure, where the $S$ and $S'$ components have different amplitudes of the superconducting gap and different spin-orbit couplings breaking horizontal and vertical mirror symmetries.
The anisotropic spin-orbit interaction generates an anomalous phase shift in the CPR that varies with the magnetic field, as explicitly shown in  Supplementary Figure\,2.
The anomalous phase is then introduced in the  CPR via a phenomenological parameter $\Gamma_B$ providing a first-order cosine component in the Fourier expansion, i.e.,  $I=\sum_n I_n \sin(\varphi) + \Gamma_B \cos(\varphi)$ (see Methods for details).
The anomalous phase $\varphi_0$ is then related to the amplitude of $\Gamma_B$. We assume a linear damping of the harmonics $I_n=I_{n,0}(1-B/B^*)$, which defines the scale parameter $B^*$.

From model II, the diode sign reversal takes place only in the presence of a sizable third harmonic component.  
Figure~\ref{highharmonics}e reports $\eta(B)$ for some values of $I_{3,0}$. By increasing $I_{3,0}$, the sign inversion gets more pronounced, whereas the maxima and minima of the rectification ($B_{max}^s \simeq 0.74 B^*$, $B_{min}^s \simeq 0.89 B^*$) are barely affected by the weight of the harmonic.
Moreover, by including more harmonics in the CPR, the lineshape of $\eta(B)$ is modified. For instance, Figure~\ref{highharmonics}f shows that a fourth-order harmonic affects the magnetic field dependence by substantially removing the sign change.\\
%\section{Discussion}

\noindent\large{\textbf{Discussion}}\normalsize\\
The comparison between our experimental findings and the proposed models reveals some important features supporting the proposed mechanisms.
In particular, for both bridges $\eta$ has an almost monotonic damping in temperature that can be explained in both models by the reduction of high-order harmonics. This is expected in long metallic weak links where the CPR evolves from highly distorted to sinusoidal-like shapes at large temperatures~\cite{lik79,gol04}.
Moreover, as shown in Figs.~\ref{shortNB}e and ~\ref{longNB}e,
$B_{max}$ is temperature resilient until $T\geq 0.5\,T_c$.
This feature is fairly captured by both models, as the maximum rectification looks almost independent of the harmonic content (see Fig.~\ref{highharmonics}b for model I and Fig.~\ref{highharmonics}e for model II). 
However, long bridges exhibit features that are mostly accounted by model I, while short ones are more compatible with model II.
For example, in long (short) bridges the sign reversal is present below (above) $B_{max}$ as shown in Fig.~\ref{longNB}b and Fig.~\ref{shortNB}b to be compared with Fig.~\ref{highharmonics}b and e, respectively. 
Multiple sign reversal nodes appear in high fields only for long bridges, as shown in Fig.~\ref{longNB}b and well described by the interferometric mechanism of model I. While the rectification lineshape given by model II presents only one inversion node.
%single inversion expected form
Moreover, the quick damping of the rectification inversion observed at low fields ($<$ 0.3\,T) in Fig.~\ref{longNB}f is captured by the vortex dynamics described in Fig.~\ref{highharmonics}c.
The relative size of the vortex $\xi/w$ is temperature-dependent and influences the vortex position in the bridge. Thus, it is plausible to expect that variations of the vortex size mostly affect the rectification lineshape 
at low fields while remaining substantially unchanged for larger fields, as shown in Fig.\ref{highharmonics}c.

Finally, it is interesting to note that by extending the proposed models to in-plane magnetic fields, a sizable supercurrent rectification is anticipated but without sign reversal. In particular, for model I, no phase shift is expected from the spatial dependence of the vector potential, since the orbital coupling between an in-plane field and the electron momentum becomes negligible. Thus, the source of phase interference with the vortex winding is eliminated. For model II, the anomalous phase and the harmonic content would be differently affected by an in-plane Zeeman field compared to the out-of-plane orientation not resulting in sign reversal.\\

%\section{Conclusions}

\noindent\large{\textbf{Conclusions}}\normalsize\\
In summary, we have demonstrated the implementation of supercurrent diodes in Nb Dayem nanobridges. By breaking the time-reversal symmetry with an out-of-plane magnetic field, we demonstrate that both the amplitude and sign of the rectification can be tuned without inverting the polarity of the applied field. The evolution of the critical current on $B$ field decays monotonically without showing any skewness, which rules out self-field effects or Meissner currents.
We have developed two theoretical models to account for the sources of time- and inversion-symmetry breaking, one based on a vortex phase winding, and one that takes into account the spin-orbit interactions present in polycrystalline heavy materials. 
Yet, a quantitative description of the supercurrent diode effect in metallic nanoconstrictions should account for both scenarios, which complement each other and can coexist.
Furthermore, the fabrication process is simple when compared to that of other platforms, a compelling step towards scalability. 
Analogous nanobridges can be realized from several elemental superconductors currently at the base of other architectures, such as nanocryotrons~\cite{mccaughan14}, rapid single-flux quanta (RSFQ)~\cite{likharev91} and memories~\cite{ligato21}, which would ease a potential integration.

Finally, the sharp sign reversal of the diode rectification allows us to envisage applications of Dayem nanobridges as $B$-field threshold detectors. When biased in the vicinity of the rectification node, small variations of an environmental magnetic field would result in modifications of the sign of the rectification parameter.\\

\noindent\large{\textbf{Methods}}\normalsize\\

\textbf{Sample fabrication.}
Nb strips and constrictions are patterned by e-beam lithography on AR-P 679.04 (PMMA) resist. PMMA residuals are removed by O$_2$-plasma etching after developing. Nb thin films were deposited by sputtering with a base pressure of 2 $\times$ 10$^{-8}$ Torr in a 4 mTorr Ar (6N purity) atmosphere and liftoff by acetone or AR-P 600.71 remover. A thin Ti layer was previously sputtered to improve Nb adhesion and base pressure in the deposition chamber, resulting in a nominal thickness Ti(4nm)/Nb(25nm) and Ti(4nm)/Nb(55nm) for the so-called short and long nanobridges, respectively.\\

\noindent\textbf{Transport measurements.}
Transport measurements were carried out in filtered (two-stage RC and $\pi$ filters) cryogen-free $^3$He-$^4$He dilution refrigerators by a standard 4-point probe technique. DC current-voltage characteristics were measured by sweeping a low-noise current bias positively and negatively, and by measuring the voltage drop across the weak links with a room-temperature, low-noise pre-amplifier for each current value every $\sim 20$\,ms. Current step size $\Delta I$ was adapted depending on the switching current keeping the values lower than $0.002 I_{sw}(B=0) \sim 1\mu$\,A  and 100\,nA for short and long devices, respectively. The switching currents are extracted from the maximum of the derivative $dV/dI$ and averaged from 5-10 reiterations of the $IV$ curves. Error bars account for the standard deviation and propagation of errors in $I_{sw}$, $\Delta I_{sw}$, and $\eta$. For the characteristic fields, the error is given by the magnetic field step $\Delta B_z <$ 10\,mT.
Joule heating is minimized by automatically switching the current off once the device turns into the normal state. 
A delay between sweeps was optimized to keep the stability of the fridge temperature lower than 50 mK. 
Furthermore, no changes in the switching currents (up to the accuracy given by the standard deviation) were observed in different cooling cycles, by changing the order of the sweeps or by adding an extra delay in the acquisition protocol concluding that hysteretic behavior or local heating is negligible.\\

%\subsection{Theoretical models}
\noindent\textbf{Theoretical models.}
The current phase relation for the model I is given by $I=\sum_{n=1,2}\int_{0}^{w} I_n \sin \left(\phi(x) +n \varphi \right) dx$. The critical current in both directions is evaluated by determining the maximum and minimum values with respect to the phase bias $\varphi$. Here, the spatial-dependent phase difference $\phi(x)$ is given by the magnetic field and the phase winding contribution due to the vortex~\cite{krasnov20}. We have assumed that the supercurrent has a subdominant second harmonic contribution, as expected in long, diffusive weak links~\cite{lik79,gol04}. 
The spatial dependence of the phase along the transverse direction $x$ is expressed as $\phi(x)=\frac{2 \pi d_b B_z x}{\Phi_0}+\phi_v(x)$ where $d_b$ is a characteristic length of the weak link, related to the width of the junction, and $\phi_v(x)=\gamma_v \arctan\left[\frac{(x-x_v)}{y_v}\right]$ is the spatially-inhomogeneous phase offset due to the vortex structure. $\gamma_v$ indicates the sign and amplitude of the winding and $(x_v,y_v)$ the position of the vortex core with respect to the boundaries (see Fig.~\ref{highharmonics}a). We also assume that the second harmonic amplitude of the supercurrent is vanishing at $B=B^{*}$.

Concerning model II, the current phase relation is given by $I=\sum_n I_n \sin(\varphi) + \Gamma_B \cos(\varphi)$. The critical currents are evaluated by determining the maximum and minimum values of the Josephson current. For the examined model, we assume that the amplitudes of the $n$-harmonic $I_n$ get suppressed by $B$ with a linear rate, i.e. $I_n=I_{n,0}(1-\frac{B}{B^*})$ for $n \geq 1$. The linear trend is compatible with the observed behavior of the overall supercurrent amplitude as shown in Fig.\,2a and 3a for the range of applied field where the rectification is nonvanishing. Furthermore, the suppression of the amplitude of the harmonics with the magnetic field can be related to the reduction of the effective transmission across the grains due to depairing and magnetic interference. The decrease in the transmission implies a reduction of the non-harmonic amplitudes~\cite{lik79,gol04}.
In this model, we have performed a real space simulation for the examined geometry. By solving the Bogoliubov-de Gennes equations on a finite-size slab in the presence of an out-of-plane magnetic field, we demonstrate that an anomalous phase shift can be obtained. The simulation is performed for a system size $N_x \times N_y$ with $N_x=150$ and $N_y=100$. The employed tight-binding model includes a nearest neighbour hopping amplitude $t$, and the conventional spin-singlet local pairing amplitude. We apply a phase bias across the weak link and determine the free energy as shown in the Supplementary Figure\,2a. The resulting anomalous phase increases with the magnetic field and depends on the strength of the Rashba and Dresselhaus interactions as shown in the Supplementary Figure\,2b. 
The linear Rashba term on a lattice for a two-dimensional geometry is expressed as $H_R=\alpha_R [\sin(k_x) \sigma_y -\sin(k_y) \sigma_x]$ while the Dresselhaus term is given by $H_D=\alpha_D [\sin(k_x) \sigma_x -\sin(k_y) \sigma_y]$ with $\sigma_i (i=x,y)$ being the Pauli matrices associated with the spin angular momentum. We notice that $H_R$ breaks the horizontal mirror symmetry while $H_D$ breaks both the vertical and horizontal mirror symmetries. This reduction of mirror symmetry is expected to be locally or globally broken in granular films~\cite{Park19}. 
Notably, the presence of the Dresselhaus term is crucial to induce an anomalous phase shift in the supercurrent in the presence of an out-of-plane magnetic field.

Concerning the size of the Rashba and Dresselhaus couplings, our analysis shows a sizable, nonvanishing anomalous phase in the presence of an out-of-plane magnetic field for values from 0.01 to 0.3, in units of the effective electron hopping amplitude $t$. We notice that $t$ is an effective electronic amplitude that describes the low energy properties of the superconductor which can be taken in the order of 20\,meV. For such value, Rashba/Dresselhaus couplings able to generate the anomalous phase will range from 0.2 to 6\,meV. Although there is no experimental estimations of Rashba and Dresselhaus coupling for Nb thin films, these values are plausible for heavy metals like Nb, since Rashba spin splitting scales with the size of the atomic spin-orbit coupling~\cite{Sunko2017,Bihlmayer2022}. %This behavior stems from the fact that 
Nb has a multi-orbital electronic structure~\cite{Russmann2022}, in this case, the breaking of the inversion or mirror symmetry primarily leads to an orbital Rashba coupling~\cite{Park2011,Mercaldo2020} among orbitals with different mirror parity. Then, the spin Rashba interaction arises through the atomic spin-orbit coupling. The large amplitude of the atomic spin-orbit coupling in Nb can thus yield a sizable spin Rashba interaction~ \cite{Sunko2017,Ishizaka2011,Yaji2010,Bihlmayer2022}.\\

%\section{Data availability}
\noindent\large{\textbf{Data availability}}\normalsize\\
\noindent \footnotesize{The data that support the findings of this study are available
from the corresponding author upon reasonable request.}\normalsize\\
%\section{Data availability}

\noindent\large{\textbf{Code availability}}\normalsize\\
\noindent \footnotesize{The code that support the findings of this study are available
from the corresponding author upon reasonable request.}\normalsize\\

\noindent\large{\textbf{References}}\normalsize\\

\vspace{1cm}
\noindent\large{\textbf{Acknowledgments}}\footnotesize\\
This work was funded by the EU’s Horizon 2020 Research and Innovation Framework Program under Grant
Agreement No. 964398 (SUPERGATE), No. 101057977 (SPECTRUM),
and by the PNRR MUR project PE0000023-NQSTI.\\

\noindent\large{\textbf{Author contributions}}\footnotesize\\
D.M. fabricated the samples, conducted the experiments, and analyzed data with inputs from A.C., E.S. and F.G. The theoretical models describing the experiment was developed by Y.F., M.T.M., and M.C. The manuscript was written by
D.M., A.C., E.S., M.C., and F.G. with inputs from all the authors. E.S. and F.G. conceived the experiment. F.G. supervised and coordinated the project. 
All authors discussed the results and their implications equally at all stages.\\

\noindent\large{\textbf{Competing interests}}\footnotesize\\
The authors declare no competing interests.\\

\noindent\large{\textbf{Additional Information}}\footnotesize\\
\textbf{Supplementary Information} The online version contains supplementary material. \\

\noindent\textbf{Correspondence} and requests for materials should be addressed to Daniel Margineda or Francesco Giazotto.
%%%%%%%%%%%%%%%%%%%%%%%%%%%%%%%%%%%%%%%%%%%%

\end{document}